\documentclass{article}

    \PassOptionsToPackage{numbers, compress}{natbib}

\usepackage[preprint]{neurips_data_2024}

\newcounter{ziyu}
\stepcounter{ziyu}

\usepackage[utf8]{inputenc} 
\usepackage[T1]{fontenc}    
\usepackage{hyperref}       
\usepackage{url}            
\usepackage{booktabs}       
\usepackage{amsfonts}       
\usepackage{nicefrac}       
\usepackage{microtype}      
\usepackage{xcolor}         
\usepackage{multirow}
\usepackage[pdftex]{graphicx}
\usepackage[skip=1pt]{caption}

\newcommand{\para}[1]{\vspace{0.1mm}\noindent\textbf{#1}.}

\title{CEBench: A Benchmarking Toolkit for the Cost-Effectiveness of LLM Pipelines}

\author{%
  Wenbo Sun \\
  Delft University of Technology\\
  \texttt{w.sun-2@tudelft.nl} \\
  \And
  Jiaqi Wang \\
  Delft University of Technology\ \\
  Sue B.V., Netherlands\\
  \texttt{j.wang-84@student.tudelft.nl} \\
  \AND
  Qiming Guo \\
  Texas A\&M University - Corpus Christi \\
  \texttt{qguo2@islander.tamucc.edu} \\
  \And
  Ziyu Li \\
  Delft University of Technology \\
  \texttt{z.li-14@tudelft.nl} \\
  \And
  Wenlu Wang \\
  Texas A\&M University - Corpus Christi \\
  \texttt{wenlu.wang@tamucc.edu} \\
    \And
  Rihan Hai \\
  Delft University of Technology \\
  \texttt{r.hai@tudelft.nl} \\
}

\begin{document}

\maketitle

\begin{abstract}
Online Large Language Model (LLM) services such as ChatGPT and Claude 3 have transformed business operations and academic research by effortlessly enabling new opportunities. However, due to data-sharing restrictions, sectors such as healthcare and finance prefer to deploy local LLM applications using costly hardware resources. This scenario requires a balance between the effectiveness advantages of LLMs and significant financial burdens. Additionally, the rapid evolution of models increases the frequency and redundancy of benchmarking efforts. Existing benchmarking toolkits, which typically focus on effectiveness, often overlook economic considerations, making their findings less applicable to practical scenarios. To address these challenges, we introduce CEBench, an open-source toolkit specifically designed for multi-objective benchmarking that focuses on the critical trade-offs between expenditure and effectiveness required for LLM deployments. CEBench allows for easy modifications through configuration files, enabling stakeholders to effectively assess and optimize these trade-offs. This strategic capability supports crucial decision-making processes aimed at maximizing effectiveness while minimizing cost impacts. By streamlining the evaluation process and emphasizing cost-effectiveness, CEBench seeks to facilitate the development of economically viable AI solutions across various industries and research fields. The code and demonstration are available in \url{https://github.com/amademicnoboday12/CEBench}.

\end{abstract}

\section{Introduction}
The rise of large language models (LLMs) has created numerous new opportunities for business owners and researchers. Many traditional areas \cite{teubner2023welcome,peng2023study,wang2024survey} have begun using LLMs to automate tasks and enhance their operations and research processes.

LLMs present impressive capabilities across tasks, with knowledge acquired during pre-training and fine-tuning \cite{roberts2020how,petroni2019language}. However, these models often face limitations due to static factual knowledge embedded within their parameters. The Retrieval-Augmented Generation (RAG) addresses this issue by integrating external knowledge bases during the generation process, which allows for access to up-to-date and domain-specific information, thereby enhancing the accuracy and relevance of the output \cite{lewis2020retrievalaugmented}. RAG continues to be a dynamic area of exploration, where recent surveys have been published that synthesize progresses and categorize specializations in integrating external knowledge to boost model performance \cite{li2022survey,gao2024retrievalaugmented}. For example, recent studies have begun to implement LLMs in the healthcare field \cite{peng2023study}, utilizing the RAG techniques to infuse domain-specific knowledge. This approach is preparing LLMs to serve as general practitioners, which could significantly reduce the workload of healthcare professionals. Furthermore, some e-commerce companies have deployed RAG-equipped LLM chatbots \cite{wei2024leveraging} to act as customer service roles. As more online LLM services emerging \cite{chatgpt,claude}, the power of LLMs becomes more accessible to general people who want to use AI to improve their business and accelerate decision-making processes.

Despite the thriving landscape of LLMs, data privacy laws such as the General Data Protection Regulation (GDPR) necessitate local data storage for specific scenarios \cite{yao2024survey}, sometimes hindering practitioners from taking advantage of cost-effective online LLM services. Consequently, they must consider deploying LLMs locally or within their private cloud infrastructures. Serving LLMs requires high-performance graphics processing units (GPUs) to maintain acceptable latency, which is resource-intensive and costly. This problem is especially critical for budget-limited research communities and small businesses. Although model compression \cite{xu2023survey} can reduce resource costs, assessing whether the consequent performance degradation is tolerable presents another challenge. Moreover, the constant emergence of new LLMs, each claiming superior performance across various tasks, complicates the model selection process for potential users. 

We identify two primary challenges that must be addressed to make optimal decisions when deploying local LLM pipelines:

\para{1. Challenge in benchmarking convenience} Despite numerous existing toolkits \cite{openaieval,harness,wu2023openicl} aimed at simplifying LLM benchmarking, significant coding efforts are still required each time an LLM application is evaluated. These tasks include model deployment, configuration of data loaders and vector databases for RAG, as well as the collection and analysis of evaluation results. These tasks require an integrated benchmarking toolkit that enables users to benchmark a wide range of LLM application scenarios with zero coding.

\para{2. Challenge in benchmarking for cost-effectiveness} Most evaluation toolkits and benchmarks prioritize the generative quality of LLMs but often neglect the cost implications of deploying these models. For instance, the Llama3 \cite{llama3} 70B model achieves a BoolQ \cite{clark2019boolq} task score of 79.0, whereas the 8B model attains 95.8\% of that performance with only 11.75\% of the memory requirement. If a slight performance degradation is acceptable, using the 8B model can significantly reduce costs, especially when combined with RAG techniques that incorporate external knowledge to improve model performance. Yet, few toolkits account for the holistic application and cost of RAG-integrated LLM pipelines, highlighting the need for benchmarks that support trade-offs in cost-effectiveness.

\para{Contributions} To address these challenges, we introduce CEBench, an open-source toolkit designed for multi-objective benchmarking with a focus on the cost and effectiveness trade-offs crucial in both business and research. CEBench's core functionalities enables users to strategically evaluate and optimize these trade-offs with simple configurations, facilitating budget-sensitive decision-making. This is particularly important for stakeholders aiming to maximize model effectiveness while minimizing cost. By streamlining the evaluation process and prioritizing cost-effectiveness, CEBench seeks to support the development of economically viable AI solutions across various industries and research fields.



\section{CEBench Toolkit}
\begin{figure}
  \centering
  \includegraphics[width=0.83\linewidth]{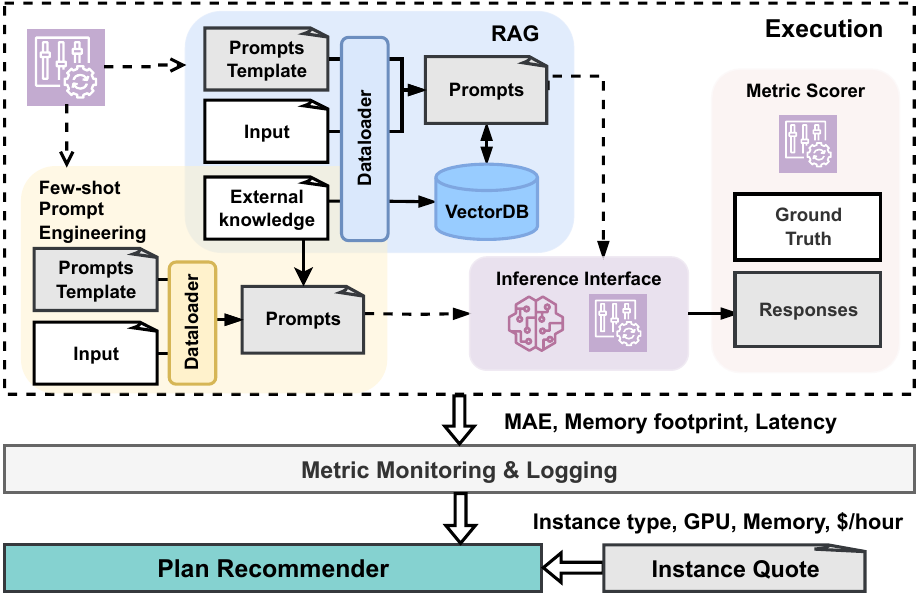}
  \caption{Workflow of benchmarking LLM pipelines using CEBench.}
  \label{fig:workflow}
  \vspace{-3mm}
\end{figure}
This section introduces the workflow of CEBench and the benchmarking scenarios it supports. Starting with an overview of the workflow and architecture, this section explains how the modules in CEBench interact and operate across various benchmarking scenarios.
\subsection{Components of CEBench}
Fig. \ref{fig:workflow} illustrates the workflow in CEBench.  The main components and their interactions are outlined below.

\para{Configuration}  CEBench uses configuration files to manage benchmark settings, including input data paths, RAG hyperparameters, LLM specifications, and evaluation metrics. These adjustments allow CEBench to automate batch experiments iteratively without coding.

\para{Dataloader} The dataloader in CEBench performs two major functions. First, it merges prompt templates and queries to generate line-separated prompt files. Second, it tokenizes and slices external knowledge into chunks, then converts these chunks into embeddings, stored in the vector database.

\para{Query execution} The query execution engine processes prompts generated by the dataloader and executes queries through the LLM inference interface. The default interface in this work is provided by the Ollama model hub\footnote{\url{https://www.ollama.com/}}. Custom models can be integrated into CEBench by implementing a customized inference interface.

\para{Metric monitoring \& logging} The performance and system resources are measured and logged. We allow both standard and customizable metrics to evaluate response quality, facilitating business-oriented benchmarking.

 \para{Plan recommender} Given all the logged metrics and budget constraints, CEBench is able to recommend optimal plans, for both cost and effectiveness, capable of illustrating a Pareto front to demonstrate optimal configurations.

The workflow in CEBench initiates batches of benchmarking tasks based on configuration files and records metrics, e.g., generative quality and memory footprint, which are crucial for multi-objective decision making. This automated execution and monitoring facilitate zero-coding benchmarking tasks, significantly improving convenience and reducing redundant coding efforts.

\subsection{Architecture Overview}
\begin{figure}
  \centering
  \includegraphics[width=0.67\linewidth]{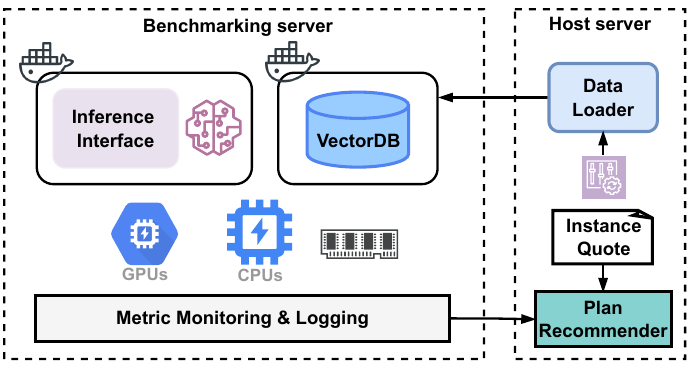}
  \caption{The architecture of CEBench. The benchmarking server and the host can be deployed on the same physical server.}
  \label{fig:arch}
  \vspace{-5mm}
\end{figure}
In addition to a detailed description of the components in CEBench, we also present an overview of the architecture in Fig. \ref{fig:arch} to illustrate how CEBench is deployed and executed. The toolkit consists of a host server managing configurations, data loading, and recommending optimal plans. The benchmarking server runs the vector database and LLMs in Docker containers. 
Furthermore, the metric logging component is deployed on the benchmarking server to monitor resource usage. 

It is important to clarify that the instance quotes do not exclusively reflect the expenses of resources used to run the benchmarks on the server.
The server for benchmarking provides an environment well-equipped with adequate resources to execute LLM pipelines. 
After collecting data on system resource usage, the plan recommender matches servers that are capable of running the LLM pipeline, and generates estimated cost based on the quotes.

\subsection{Benchmarking Scenarios in CEBench}
In this section, we explore various benchmarking scenarios supported by CEBench. Each scenario is designed to address specific needs of deploying and evaluating LLM pipelines, ensuring that users can comprehensively assess both generative quality and cost-effectiveness. As an integrated toolkit, CEBench simplifies the benchmarking process across multiple dimensions, including model generative quality, resource utilization, and financial considerations.

\subsubsection{Benchmarking effectiveness of LLMs}

CEBench facilitates the benchmarking of various LLMs by providing a comprehensive framework that evaluates their performance based on predefined metrics. This scenario involves configuring the LLMs with different settings and assessing their generative capabilities. The process includes loading models, configuring inputs by specifying prompt templates and actual prompts to test different aspects of LLM performance, and evaluating the models based on metrics such as accuracy, fluency, coherence, and relevance of generated responses. While CEBench primarily focuses on benchmarking scenarios for locally deployed LLMs, it also provides interfaces enabling evaluation of online LLMs such as ChatGPT (GPT-4) \cite{chatgpt} and Claude 3 \cite{claude}. In Section \ref{sec:contract}, we will demonstrate how to utilize our benchmarking toolkit to evaluate budgets with online LLMs.

\subsubsection{End-to-End benchmarking for RAG pipelines}

This scenario extends the benchmarking process to include  RAG techniques. CEBench evaluates the performance of LLMs when combined with external knowledge sources. The steps include integrating LLMs with vector databases to facilitate RAG, configuring external knowledge bases, and tuning parameters such as retrieval algorithms and top K documents to be considered. The system assesses the end-to-end performance, including the accuracy of retrieved information and the quality of the final output generated by the LLM, while also measuring the resource consumption and cost-effectiveness of the combined LLM and RAG setup.

\subsubsection{Benchmarking prompt engineering}

CEBench allows users to benchmark different prompting techniques \cite{marvin2023prompt} to optimize LLM performance. This scenario involves creating and modifying prompt templates to test their impact on LLM responses, evaluating LLM performance in zero-shot, few-shot, and long-context scenarios to understand the effectiveness of various prompts in generating responses.



\subsubsection{Multi-objective evaluation}

This scenario involves benchmarks regarding multiple objectives, including generative quality, time consumption, and financial cost, assessing the trade-offs between multiple objectives to identify the most balanced configuration, and providing deployment recommendations for stakeholders to make informed decisions about deploying LLMs and RAG pipelines.

When benchmarking locally deployed LLMs, CEBench tracks memory usage and processing latency. Memory usage indicates the minimal memory requirement for candidate server instances, while latency helps estimate the costs of running LLM pipelines on untested hardware resources. Specifically, the instance quote provides the compute performance of GPUs, measured in teraflops. By comparing the performance ratio between the target GPU and the GPU used in the benchmark server, we can estimate the latency of the tested pipelines. This estimated latency can further aid in calculating the costs of processing a similar volume of prompts.

\section{Comparison with Existing Benchmarking Toolkits}
In this section, we compare CEBench with existing LLM benchmarking toolkits. This comparison falls into two categories. The first category includes the skeleton frameworks for constructing LLM applications that can serve as benchmarking toolkit, such as LangChain \cite{LangChain}, LlamaIndex \cite{LlamaIndex}, and Flowise \cite{flowise}. The second category consists of specialized LLM benchmarking toolkits, including OpenAI Evals \cite{openaieval}, Harness \cite{harness}, OpenICL \cite{wu2023openicl}, and LLMeBench \cite{dalvi2023llmebench}.
\begin{table}[h]
\vspace{-4mm}
\caption{Functionality comparison between CEBench and various popular benchmarking toolkits.}
\label{tab:comparison}
\resizebox{\textwidth}{!}{%
\begin{tabular}{l|cccccccc}
\hline
Functionalities\textbackslash{}Toolkits & \multicolumn{1}{l}{\textbf{CEBench}} & \multicolumn{1}{l}{LangChain} & \multicolumn{1}{l}{LlamaIndex} & \multicolumn{1}{l}{Flowise} & \multicolumn{1}{l}{\begin{tabular}[c]{@{}l@{}}OpenAI \\ Evals\end{tabular}} & \multicolumn{1}{l}{Harness} & \multicolumn{1}{l}{LLMeBench} & \multicolumn{1}{l}{OpenICL} \\ \hline
Customized data & \checkmark & \checkmark & \checkmark & \checkmark &  & \checkmark & \checkmark & \checkmark \\
Local LLMs & \checkmark & \checkmark & \checkmark & \checkmark & \checkmark & \checkmark & \checkmark & \checkmark \\
RAG integrated & \checkmark & \checkmark & \checkmark & \checkmark &  &  & \checkmark &  \\
Zero-code & \checkmark &  &  & \checkmark &  &  &  &  \\
Budget constraint & \checkmark &  &  &  &  &  &  &  \\
Plan recommendation & \checkmark &  &  &  &  &  &  &  \\ \bottomrule
\end{tabular}%
}
\vspace{-4mm}
\end{table}

Table \ref{tab:comparison} demonstrates the functionalities supported by the various benchmarking toolkits discussed. We observe that open-source skeleton frameworks generally offer greater flexibility to support various modules, such as customized data input and RAG. Notably, Flowise supports graphical pipeline compilation without coding efforts.

In contrast, specialized frameworks usually serve specific purposes. For instance, OpenICL and Harness only support local LLMs with predefined prompts as inputs. OpenAI Evals is a simple toolkit designed to validate the capabilities of LLMs using widely adopted public benchmarking tasks.

None of the existing toolkits are compatible given budget constraints. However, with minor code changes, these frameworks can be adapted to monitor system resource usage. Nevertheless, the metrics and design objectives of these frameworks are not intended to generate optimal deployment plans. This distinction highlights the unique value of CEBench in benchmarking tasks for cost-effectiveness.



\section{Use Cases for Demonstration}
This section introduces two use cases demonstrating how CEBench determines the best resource configurations for LLM pipelines. The first involves using an LLM to score psychological questionnaires, requiring local storage of clinical data and the LLM due to data-sharing restrictions. The second involves labeling legal documents, permitting the use of online LLM services. These examples illustrate how CEBench helps practitioners achieve an optimal balance of cost and effectiveness for both locally deployed LLM pipelines and online LLM services. 

\subsection{Use Case 1: Mental Health LLM Assistant}
Mental health issues affect nearly a billion people globally, including 14\% of adolescents as reported by the World Health Organization in 2019 \cite{WHOmental}. The advancement of LLMs offers promising solutions in mental health fields such as prevention, surveys, pre-diagnosis, and treatment assistance, especially where medical resources are limited. Normal mental health pre-diagnosis uses dialogues for preliminary assessments, requiring highly responsive models that understand and address mental health concerns accurately. Given strict data-sharing regulations, local deployment of LLM applications is necessary, although it incurs significant hardware maintenance costs. Thus, benchmarking to identify optimal model configurations is essential for cost-effectiveness. Details on data-sharing and maintenance costs are discussed in the Appendix \ref{app:mental}.

\subsubsection{Dataset: DAIC-WOZ}
The DAIC-WOZ database~\cite{gratch2014distress}, established in 2014, comprises 187 dialogues averaging 16 minutes each, aimed at recognizing signs of mental illness through verbal and nonverbal cues. Data is categorized into folders by patient IDs and includes transcripts, audio, and video files, along with key metrics like PHQ-8 scores and participants' demographic details, offering insights into each participant’s mental health. The detailed data description and data preprocessing can be found in Appendix \ref{app:mental}.

\subsubsection{Benchmarking configurations}
\begin{table}[t]
\caption{Configurations in the pipeline of mental health LLM assistant.}
\label{tab:var_1}
\centering
\resizebox{\textwidth}{!}{%
\begin{tabular}{l|l|l}
\toprule
\textbf{Variables}    & \textbf{Values}                                                                                                                     & \textbf{Description}                                               \\ \hline
Quantization & \begin{tabular}[c]{@{}l@{}}no-quantization (no),\\ scalar-quantization (sq),\\ product-quantization (pq)\end{tabular}                                  & Quantization method for the embeddings                    \\ \hline
Top K        & 2, 5, 10                                                                                                                   & Top K of the results retrieved from vector database       \\ \hline
Chunck Size  & 500, 1000, 2000                                                                                                            & Number of Characters in each chunk in the vector database \\ \hline
LLM          & \begin{tabular}[c]{@{}l@{}}llama3:8b\_{[}4bit, 8bit, 16bit{]},\\ llama3:80b\_4bit,\\ mixtral:8x {[}7b, 22b{]}\end{tabular} & LLM used for mental health scoring                      \\ \bottomrule 
\end{tabular}%
}
\vspace{-4mm}
\end{table}

\begin{table}[t]
\caption{Quote of AWS instances and the benchmarking server. TFLOPs is a measurement of attainable performance of GPUs.}
\label{tab:quote}
\small
\centering
\begin{tabular}{l|l|l|l|l}
\toprule
\textbf{Instance type}                                                 & \textbf{GPU}             & \textbf{GPU memory} & \textbf{TFLOPs} & \textbf{\$/hour} \\ \midrule
EC2 P2                                                        & K80             & 12GB       & 8.22   & 1.326   \\ 
EC2 P3                                                        & V100            & 16GB       & 32.71  & 3.823   \\ 
EC2 G5                                                        & A10G            & 24GB       & 31.52  & 1.515   \\ 
EC2 G6                                                        & L4 Tensor Cores & 16GB       & 30.29  & 1.172   \\ \midrule
\begin{tabular}[c]{@{}l@{}}Benchmarking\\ Server\end{tabular} & A100            & 80GB       & 77.97  & 4.777       \\ \bottomrule
\end{tabular}
\vspace{-4mm}
\end{table}
\label{eval:mental}
Local deployment of LLM applications involves numerous variables that can significantly impact cost and effectiveness. To comprehensively benchmark local LLMs with a RAG pipeline, we evaluate multiple openly available LLMs and various vector database settings, ensuring a thorough assessment of how different configurations influence effectiveness. The configurations are detailed in Table \ref{tab:var_1}.

The benchmarking server is equipped with an Nvidia A100 GPU, though this setup is not strictly required for LLM applications. CEBench tracks resource usage, such as memory footprint, and records underutilization to identify suitable hardware settings from our instance quote. For this, we use AWS EC2 instance configurations and prices\footnote{\url{https://aws.amazon.com/ec2/instance-types/}} as examples, as detailed in Table \ref{tab:quote}.

We set three objectives for this use case: \texttt{i)} \emph{End-to-End Latency} (measured in seconds) measures the time consumption of local LLMs with RAG, which is crucial since locally deployed models are usually slower than online services. \texttt{ii)} \emph{Mean Absolute Error} (MAE) measures the quality of responses. The input questionnaires have integer scores, and the gap between predictions and labels is used to assess accuracy. \texttt{iii)} \emph{Cost} is assessed by estimating the cost-effectiveness of LLM pipelines. CEBench estimates costs for various instance types by multiplying the per-hour cost of each instance by the estimated latency for processing 1,000 prompts, using the ratio of teraflop operations between the GPU on the respective instances and that on the benchmarking server.

\subsubsection{Results}

\begin{figure}[t]
  \centering
  \includegraphics[width=\linewidth]{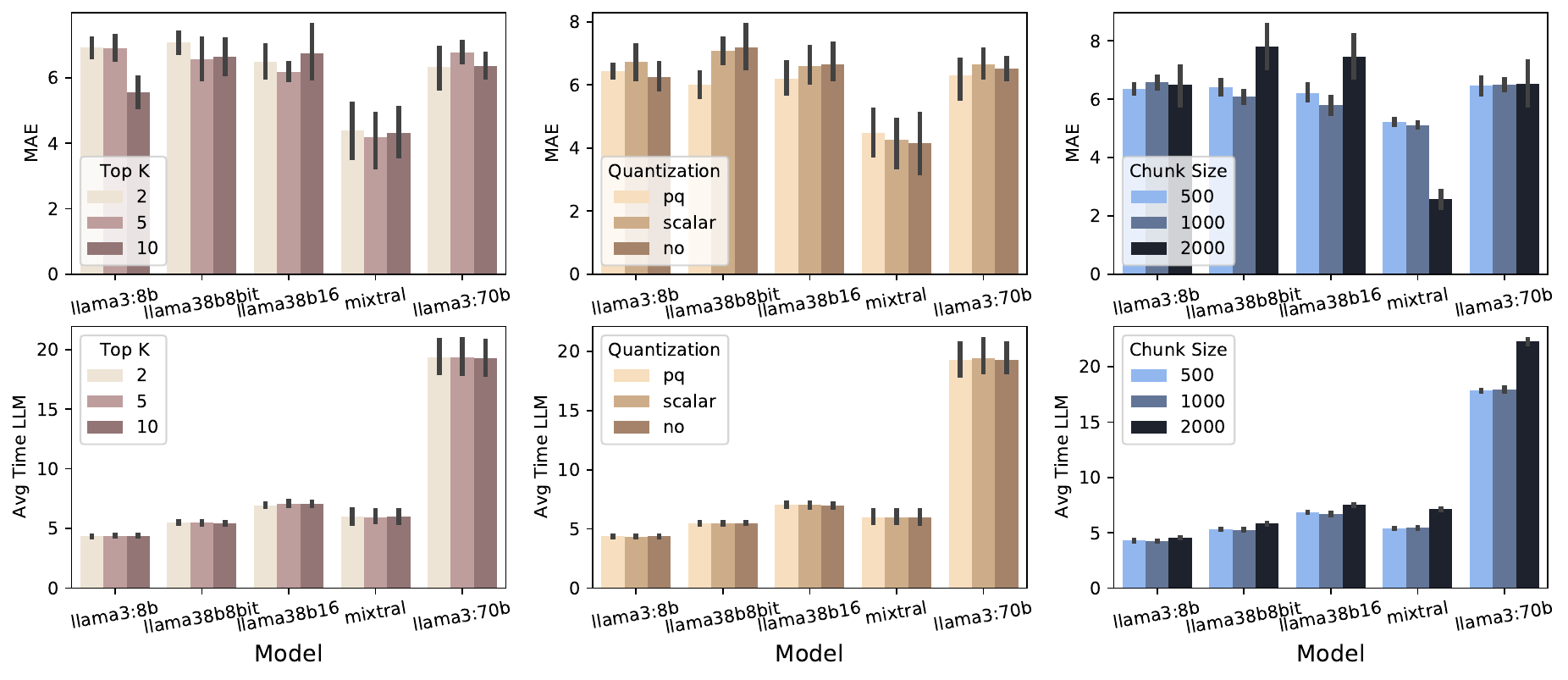}
  \caption{Average MAE and end-to-end latency regarding three variables. Mixtral:8x7b gives more accurate scores; llama3:8b has the lowest latency. }
  \label{fig:mental_all}
  \vspace{-3mm}
\end{figure}
\begin{figure}[t]
  \centering
  \includegraphics[width=\linewidth]{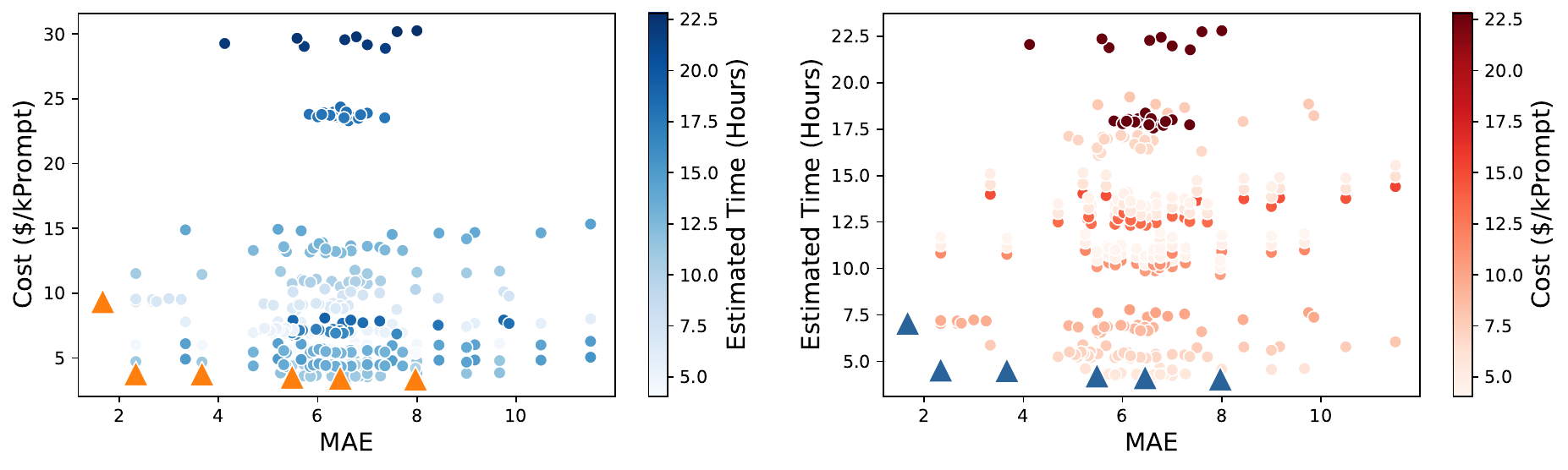}
  \caption{Trade-offs between cost, MAE and estimated time. Each sample in plot represents a LLM pipeline running on an instance. The triangle marker means the non-dominated solutions.}
  \label{fig:mental_pareto}
  \vspace{-4mm}
\end{figure}
Fig. \ref{fig:mental_all} illustrates the MAE and the end-to-end latency of LLM pipelines for various variables. mixtral:8x7b generally provides more accurate mental health scores compared to other models. The model llama3:8b exhibits the lowest latency, attributed to its smaller size. Notably, the results exclude mixtral:7x22b due to out-of-memory errors encountered during prompt loading. Despite its larger size, llama3:70b does not outperform the smaller models, yet it incurs the highest latency.

Upon examining the variables associated with the vector database, it becomes evident that the quantization method does not significantly impact the efficiency or effectiveness. The parameters `Top K' and `chunk size' have varying effects across different models. Specifically, a larger chunk size enhances the effectiveness of mixtral but negatively impacts llama3:8b. In contrast, llama3:8b benefits from retrieving fewer results from the vector database.

Fig. \ref{fig:mental_pareto} illustrates the performance objectives achieved by combinations of LLM pipelines and instances as listed in Table \ref{tab:quote}. The majority of instances incur costs lower than \$15 per 1,000 prompts. Within the set of non-dominant instances, a cluster of samples is concentrated at the lower end of the cost spectrum, indicating minimal cost variations but significant differences in MAE. Consequently, plan enumerator can recommend a set of instances with LLM pipelins, as listed in Table \ref{tab:pareto}. If higher accuracy in scoring is crucial, instance 6 is recommended. Conversely, for a more cost-sensitive application, instance 4 offers a slight reduction in accuracy but at a lower cost. Similar decision process can be applied in trade-offs between latency and MAE as shown in the right figure.

In this use case, we demonstrate that CEBench is capable of facilitating trade-offs between cost and effectiveness through comprehensive benchmarks and resource usage estimation.

\begin{table}[t]
\caption{Pareto-front of trade-offs between cost and MAE.}
\label{tab:pareto}
\small
\centering
\resizebox{\textwidth}{!}{%
\begin{tabular}{cc|ccccccc}
\hline
Id &Model     & MAE  & Est. Time (s) & Instance & Top K & Quantization & Chunk Size & Est. Cost (\$/kPrompt) \\ \toprule
1 &llama3:8b & 6.45 & 10.65        & G6       & 2     & sq       & 1000 &3.47                  \\ 
2& llama3:8b & 7.89 & 10.44        & G6       & 5     & pq    &1000       & 3.39                  \\ 
3 &llama3:8b & 5.48 & 10.89        & G6       & 5     & no  &1000         & 3.54                  \\ 
4& \underline{llama3:8b} & \underline{2.33} & 11.68        & G6       & 10    & sq  &2000     & \underline{3.80}                  \\ 
5 &llama3:8b & 3.67 & 11.61        & G6       & 10    & no   &2000        & 3.79                  \\ 
6 & \underline{mixtral:8x7b}   & \underline{1.67} & 7.06         & A100     & 5     & no &2000          & \underline{9.37}                  \\ \bottomrule
\end{tabular}
}
\vspace{-4mm}
\end{table}

\subsection{Use Case 2: Contract Reviewing}
\label{sec:contract}


The legal domain demands a sophisticated benchmarking to evaluate language models' comprehension of complex legal documents. In an international context, LLM pipelines must integrate changes in laws and legal precedents from various countries.

For contract review simulations, we use RAG and few-shot prompting to adapt models to legal changes and case law. Business owners prefer online LLM services to avoid maintenance costs of high-performance servers, although contract lengths and additional text import costs must be considered. This use case demonstrates how CEBench determines the best resource setup for RAG-equipped online LLM services to balance accuracy and cost

\subsubsection{Dataset: ContractNLI}
For contract reviewing tasks in the legal field, we use the ContractNLI dataset \cite{koreeda2021contractnli} as a benchmark. The ContractNLI dataset is designed for document-level natural language inference (NLI) on contracts, aiming to automate or support the time-consuming process of contract review. In this task, the system is presented with a set of hypotheses (e.g., "Some obligations of Agreement may survive termination.") and a contract. It must determine whether each hypothesis is \emph{entailed by}, \emph{contradicts}, or is not mentioned (\emph{neutral}) in the contract. Additionally, the system must identify evidence for its decision as specific spans within the contract. The detailed data preprocessing and description can be found in Appendix \ref{app:nli}.


\subsubsection{Benchmarking configurations}


To integrate external knowledge with LLM services, we benchmark two methods in this use case: few-shot prompting and RAG. Few-shot prompting uses five examples extracted from the training set. For the RAG method, the entire training set is stored in the vector database, which uses default settings due to the assumed absence of specialized engineers. Online LLM service options include Haiku, Sonnet, Opus from Claude 3 \cite{claude}, and GPT-4 \cite{chatgpt} from OpenAI, each differing in cost and effectiveness.

Performance is assessed using three metrics: F1 score for effectiveness, end-to-end latency (Time), and cost per prompt. The goal is to find the most cost-effective LLM service that maintains over 90\% F1 score.

\subsubsection{Results}

\begin{figure}[t]
  \centering
  \includegraphics[width=\linewidth]{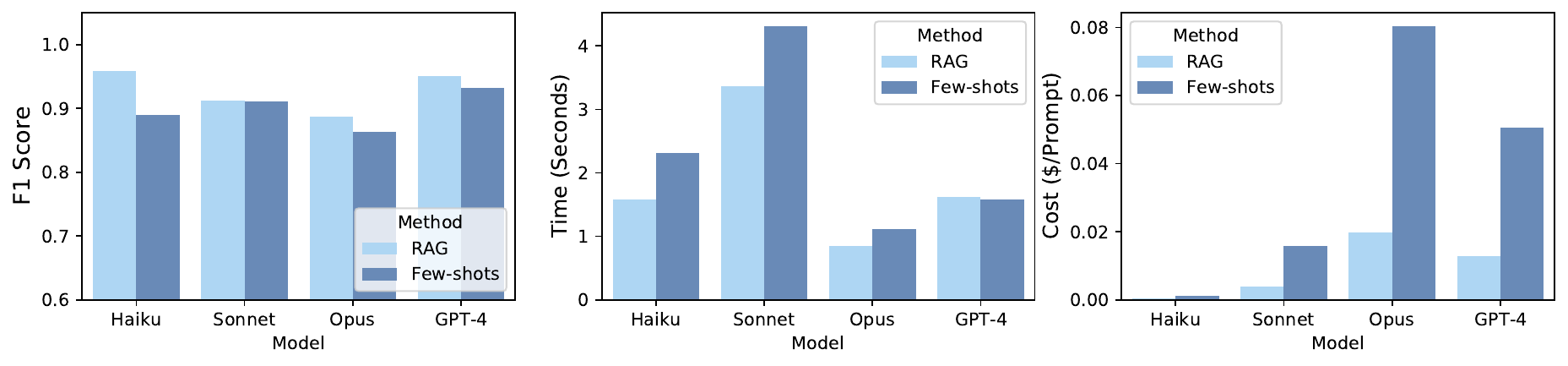}
  \caption{Results of the contract review task using online LLM services. RAG can generally improve the generative quality and reduce the cost.}
  \label{fig:contract}
  \vspace{-4mm}
\end{figure}

The results presented in Fig. \ref{fig:contract} indicate that Haiku with RAG, Sonnet, and GPT-4 meet the F1 score constraint of greater than 90\%. Haiku, the most lightweight model in the Claude 3 series, achieves the highest accuracy with the assistance of RAG. All models with RAG outperform the few-shot prompting pipelines in terms of accuracy, underscoring the effectiveness of RAG in contract review tasks. Furthermore, the latency results reveal that pipelines with RAG generally exhibit lower end-to-end latency compared to those with few-shot prompting, thanks to the shorter chunks retrieved from the vector database.

The cost analysis shows the average cost per prompt in the test dataset, with Haiku demonstrating a significantly lower cost than other pipelines. Notably, the cost per prompt for Haiku with RAG is only \$0.0003, attributed to the shorter input tokens in the RAG pipeline. Specifically, the average input token length for RAG pipelines is 1276, while the few-shot prompting pipeline averages 5039 tokens, reflecting the cost differences per prompt. When considering the superior effectiveness indicated by the F1 score, it is evident that the Haiku model with the RAG pipeline offers the best cost-effectiveness trade-offs.

This benchmarking case illustrates CEBench's capability to identify the most cost-effective LLM pipelines, particularly those involving online LLM services. Combining with the results in Section~\ref{eval:mental}, we find that the cost of online LLM services can be three times cheaper than locally deployed LLM pipelines. As computing power becomes more affordable, online LLM services will become increasingly cost-effective, further enhancing the viability of AI to the general public.

\section{Conclusion}
The rise of LLMs has transformed business decision-making and research process but also introduces challenges in selecting cost-effective deployment plans due to substantial resource requirements. We propose CEBench, an open-source LLM benchmarking toolkit, to address redundant benchmarking efforts and financial concerns. CEBench can launch batch benchmarking tasks from configuration files, thereby reducing coding efforts and improving benchmarking convenience. Moreover, CEBench supports customized LLMs and prompts, facilitating the benchmarking of generative quality for LLM pipelines involving RAG, few-shot prompting, and long context. For those with financial concerns, the resource usage monitor in CEBench enables cost estimation on alternative hardware, further allowing users to make multi-objective decisions regarding generative quality and cost. With these features, CEBench is capable of finding economically viable AI solutions for both business and research communities.

\para{Limitations} The latency estimation in this work is based on the compute performance of GPUs and is not sufficiently accurate. In the future, we plan to integrate more sophisticated performance models to better predict the latency of LLMs on untested hardware, providing more precise cost estimates.

\section*{Acknowledgment}
This work was partially supported by the US Department of Transportation (USDOT) Tier-1 University Transportation Center (UTC) Transportation Cybersecurity Center for Advanced Research and Education (CYBER-CARE). (Grant No. 69A3552348332).


\bibliography{references}
\bibliographystyle{abbrvnat}





\section*{Checklist}

The checklist follows the references.  Please
read the checklist guidelines carefully for information on how to answer these
questions.  For each question, change the default \answerTODO{} to \answerYes{},
\answerNo{}, or \answerNA{}.  You are strongly encouraged to include a {\bf
justification to your answer}, either by referencing the appropriate section of
your paper or providing a brief inline description.  For example:
\begin{itemize}
  \item Did you include the license to the code and datasets? \answerYes{See Section~\ref{gen_inst}.}
  \item Did you include the license to the code and datasets? \answerNo{The code and the data are proprietary.}
  \item Did you include the license to the code and datasets? \answerNA{}
\end{itemize}
Please do not modify the questions and only use the provided macros for your
answers.  Note that the Checklist section does not count towards the page
limit.  In your paper, please delete this instructions block and only keep the
Checklist section heading above along with the questions/answers below.

\begin{enumerate}

\item For all authors...
\begin{enumerate}
  \item Do the main claims made in the abstract and introduction accurately reflect the paper's contributions and scope?
    \answerYes{}
  \item Did you describe the limitations of your work?
    \answerYes{We have discussed limitations in Section 5}
  \item Did you discuss any potential negative societal impacts of your work?
    \answerNo{As a benchmarking toolkit, it does not create any data and models that may cause negative societal impacts.}
  \item Have you read the ethics review guidelines and ensured that your paper conforms to them?
    \answerYes{}
\end{enumerate}

\item If you are including theoretical results...
\begin{enumerate}
  \item Did you state the full set of assumptions of all theoretical results?
    \answerNA{}
	\item Did you include complete proofs of all theoretical results?
    \answerNA{}
\end{enumerate}

\item If you ran experiments (e.g. for benchmarks)...
\begin{enumerate}
  \item Did you include the code, data, and instructions needed to reproduce the main experimental results (either in the supplemental material or as a URL)?
    \answerYes{The code and reproducing instructions are included in the public repository. The link has been provided in Abstract.}
  \item Did you specify all the training details (e.g., data splits, hyperparameters, how they were chosen)?
    \answerNA{}
	\item Did you report error bars (e.g., with respect to the random seed after running experiments multiple times)?
    \answerYes{We run experiments multiple times. Results are in Section 4 and Appendix.}
	\item Did you include the total amount of compute and the type of resources used (e.g., type of GPUs, internal cluster, or cloud provider)?
    \answerYes{The hardware configurations are presented in Table 3.}
\end{enumerate}

\item If you are using existing assets (e.g., code, data, models) or curating/releasing new assets...
\begin{enumerate}
  \item If your work uses existing assets, did you cite the creators?
    \answerYes{The dataset we use for demonstration are public dataset. Relevant literature are cited.}
  \item Did you mention the license of the assets?
    \answerNo{All resources used in this work are open-source resources.}
  \item Did you include any new assets either in the supplemental material or as a URL?
    \answerYes{We have modified the ContractNLI dataset to adapt to few-shot prompting. We have included the preprocessing script in the code repository.}
  \item Did you discuss whether and how consent was obtained from people whose data you're using/curating?
    \answerNA{}
  \item Did you discuss whether the data you are using/curating contains personally identifiable information or offensive content?
    \answerNo{Anonymization has been done by the dataset providers.}
\end{enumerate}

\item If you used crowdsourcing or conducted research with human subjects...
\begin{enumerate}
  \item Did you include the full text of instructions given to participants and screenshots, if applicable?
    \answerNA{}
  \item Did you describe any potential participant risks, with links to Institutional Review Board (IRB) approvals, if applicable?
    \answerNA{}
  \item Did you include the estimated hourly wage paid to participants and the total amount spent on participant compensation?
    \answerNA{}
\end{enumerate}

\end{enumerate}


\appendix

\newpage
\section{Appendix}

\subsection{ContractNLI}
\label{app:nli}
For contract reviewing tasks in the legal field, we use the ContractNLI dataset \cite{koreeda2021contractnli} as a benchmark. The ContractNLI dataset is designed for document-level natural language inference (NLI) on contracts, aiming to automate or support the time-consuming process of contract review. In this task, the system is presented with a set of hypotheses (e.g., "Some obligations of Agreement may survive termination.") and a contract. It must determine whether each hypothesis is entailed by, contradicts, or is not mentioned (neutral) in the contract. Additionally, the system must identify evidence for its decision as specific spans within the contract.

Compared to LegalBench \cite{guha2024legalbench}, our approach not only utilizes the corpus but also includes the surrounding text to provide more context for the judgement of language models. This additional context aims to enhance the models' understanding and improve their decision-making capabilities.
One Example would be like in the Figure \ref{fig:contract-nli-example}:

\begin{figure}[h]
    \centering
    \includegraphics[width=\linewidth]{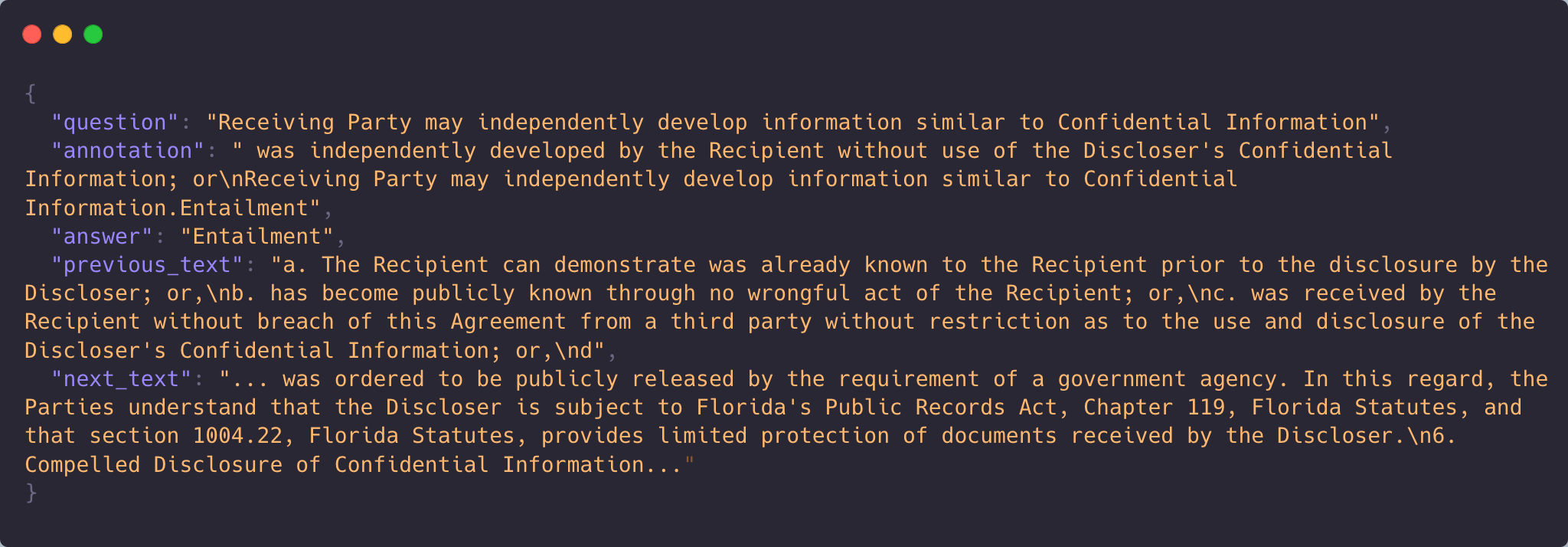}
    \caption{Example of Contract NLI questions.}
    \label{fig:contract-nli-example}
\end{figure}

\subsubsection{\textbf{Data Processing}}

In our experiments, we exclusively use the test set to evaluate the model's performance. The training sets are stored in a vector store, forming the knowledge base for the Retrieval-Augmented Generation (RAG) system. Given the limitations of the context window (models before 2023 typically have a context window of 8192 tokens or fewer), we restrict the number of tokens to 1000 for each question. This ensures that models with smaller context windows can still handle five-shot learning plus one test question per trial.

\subsubsection{\textbf{Dataset Analysis}}

To better understand the dataset, we analyze the distribution of span lengths within the ContractNLI dataset. The span lengths represent the lengths of contract excerpts used for the evaluation. Understanding this distribution helps in designing more effective experiments and provides insights into the dataset's complexity.

\begin{figure}
    \centering
    \caption{Histogram of span lengths in the ContractNLI dataset.}
    \includegraphics[width=\linewidth]{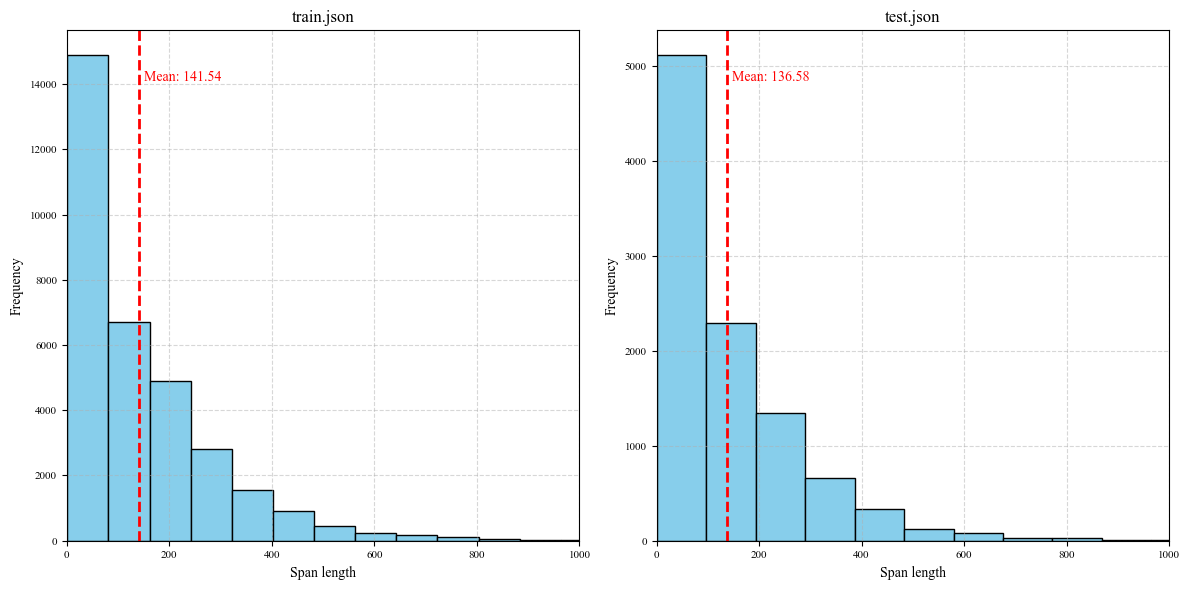}
    \label{fig:contractnli-hist}
\end{figure}

The distribution of span lengths is illustrated in Figure \ref{fig:contractnli-hist}. This histogram shows the frequency of various span lengths, with a clear indication of the mean span length.

\begin{figure}
    \centering
    \includegraphics[width=0.8\linewidth]{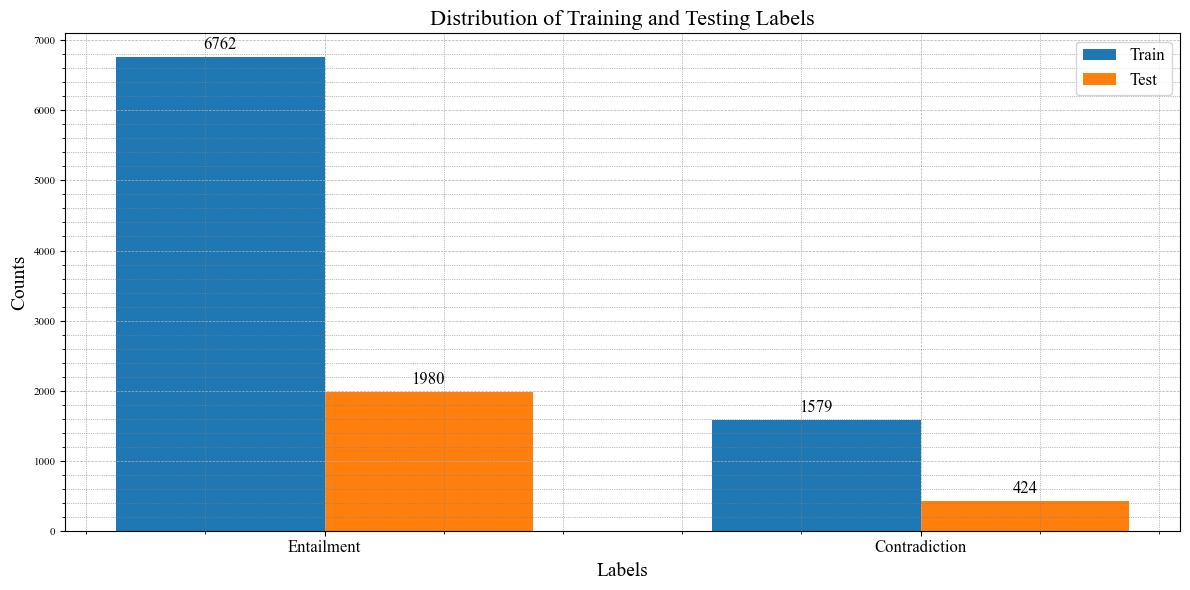}
    \caption{Label distributions of Contract NLI.}
    \label{fig:contractnli-label-dist}
\end{figure}

\subsubsection{\textbf{Experimental Setup}}

We implemented two distinct pipelines to compare their performance across different dimensions. The first was a RAG pipeline, which retrieves relevant document chunks from a vector database based on the similarity to the given question. The second was a few-shot pipeline, which utilizes the first five questions as in-context examples to prime the model for chain-of-thought reasoning.
We evaluated these pipelines using a diverse set of 400 questions, encompassing a broad range of models. These included commercial offerings such as the Claude 3 family and GPT-4, as well as locally deployed models like LLaMa 2 and LLaMa 3 families.

\subsubsection{\textbf{Results and Discussion}}
\begin{table}[h]
\centering
\caption{Performance Metrics and Average Inference Times for Different Models on Contract NLI.}
\label{tab:results-model}
\begin{tabular}{@{}cccccc@{}}
\toprule
                       &           & Haiku         & Sonnet        & Opus          & GPT-4         \\ \midrule
\multicolumn{1}{|c|}{\multirow{2}{*}{F1 Score}} &
  \multicolumn{1}{c|}{RAG} &
  \multicolumn{1}{c|}{0.9585} &
  \multicolumn{1}{c|}{0.9125} &
  \multicolumn{1}{c|}{0.8865} &
  \multicolumn{1}{c|}{0.951} \\ \cmidrule(l){2-6} 
\multicolumn{1}{|c|}{} &
  \multicolumn{1}{c|}{Few-shots} &
  \multicolumn{1}{c|}{0.8895} &
  \multicolumn{1}{c|}{0.9115} &
  \multicolumn{1}{c|}{0.863} &
  \multicolumn{1}{c|}{0.9325} \\ \midrule
\multicolumn{1}{|c|}{\multirow{2}{*}{Time Avg (s)}} &
  \multicolumn{1}{c|}{RAG} &
  \multicolumn{1}{c|}{1.585 ± 0.970} &
  \multicolumn{1}{c|}{3.366 ± 1.949} &
  \multicolumn{1}{c|}{0.847 ± 0.453} &
  \multicolumn{1}{c|}{1.612 ± 2.090} \\ \cmidrule(l){2-6} 
\multicolumn{1}{|c|}{} & Few-shots & 2.318 ± 2.318 & 4.307 ± 2.608 & 1.109 ± 0.430 & 1.582 ± 1.528 \\ \bottomrule
\end{tabular}%
\end{table}

The performance and computational efficiency of various models were evaluated on the Contract NLI task, with results summarized in Table \ref{tab:results-model}. For the RAG pipeline, the Haiku model demonstrated superior performance, achieving an F1 score of 0.9585, outperforming other models like Sonnet (0.9125), Opus (0.8865), and GPT-4 (0.951). However, when transitioning to the Few-shots pipeline, which assesses a model's ability to adapt quickly to new domains with limited data, the performance rankings shifted. GPT-4 emerged as the top-performing model with an F1 score of 0.9325, while all other models experienced a decline in scores compared to their RAG counterparts.

Regarding inference times, a crucial factor in real-world deployments, Haiku exhibited remarkable efficiency, recording the lowest average times for both RAG (1.585 ± 0.970 s) and Few-shots (2.318 ± 2.318 s) pipelines. As expected, larger models like Sonnet and Opus incurred higher inference costs, with Sonnet being the slowest for RAG (3.366 ± 1.949 s) and Opus for Few-shots (1.109 ± 0.430 s). Interestingly, despite its considerable size, GPT-4 maintained competitive inference times, clocking in at 1.612 ± 2.090 s for RAG and 1.582 ± 1.528 s for Few-shots, highlighting the optimizations in its architecture.

One example

\begin{itemize}
    \item Based on the information provided, it's unclear whether the hypothesis ""Confidential Information shall only include technical information"" is entailed or contradicted. The text mentions that Confidential Information is being provided by the Broker regarding businesses for sale, but does not specify what types of information are included. Without more context about the scope of Confidential Information, there is not enough information to determine if the hypothesis is supported or refuted by the text
\end{itemize}

\subsubsection{\textbf{Use case}}

The legal domain presents a unique set of challenges that necessitate a carefully designed benchmark framework for evaluating language models' performance. One of the primary challenges lies in the long and intricate contexts prevalent in legal documents, which rigorously test a model's understanding and inferencing abilities. Unlike concise texts, the extensive nature of legal contexts demands a high level of comprehension and reasoning skills from the model to accurately capture the nuances and implications \cite{martin2024better}.

In the legal realm, clients and practitioners alike place a premium on accuracy and explainability. The RAG and Few-shots pipelines employed in this benchmark framework cater to these critical requirements in distinct yet complementary ways. The RAG pipeline leverages external knowledge sources, enabling the model to draw upon relevant information and provide contextually accurate responses. Conversely, the Few-shots pipeline emphasizes the model's ability to rapidly adapt to new domains and tasks with limited data, offering transparency and explainability in its decision-making process.

Moreover, the legal domain often demands the identification of ambiguities, loopholes, and potential areas of contention within extensive textual contexts. The benchmark framework's emphasis on long-form understanding and inferencing capabilities directly addresses this need, allowing models to scrutinize intricate legal documents and uncover potential ambiguities or inconsistencies that may have significant implications.

Crucially, our benchmark framework focuses on the trade-off between effectiveness and efficiency, a critical consideration in the legal field where time, costs, and resource utilization are paramount concerns. By evaluating models' performance metrics alongside their computational requirements and inference times, the benchmark provides insights into the balance between accuracy and the associated operational overheads. This approach ensures that the selected language models not only deliver reliable legal insights but also align with practical considerations, such as energy consumption, computational costs, and time constraints, which are essential for real-world deployment in legal settings.

\subsection{Use Case for a Mental Health LLM Assistant}
\label{app:mental}
\subsubsection{Background}

Mental health issues are a major global concern. The 2019 World Health Organization report highlights that nearly a billion people, including 14\% of adolescents worldwide, suffer from mental disorders \cite{WHOmental}. These disorders are the primary cause of disability. Moreover, they significantly reduce life expectancy by 10 to 20 years due to preventable physical illnesses. Factors like childhood abuse, bullying, socioeconomic disparities, along with global crises such as pandemics and environmental challenges, have exacerbated mental health issues, causing a more than 25\% increase in depression and anxiety during the first year of the pandemic. Recognizing the severe impacts of mental disorders on individuals and society, the WHO urges nations to enhance mental health services and support.

In response, as Large Language Models (LLMs) advance, deploying them in various aspects of mental health care—including prevention, surveys, initial diagnosis, and treatment assistance—emerges as a promising approach. This strategy offers a viable solution for addressing the challenge of providing adequate mental health services to a large population, particularly in resource-limited settings.

\subsubsection{Task Setting}

Exploring the optimal model and settings for specific mental health tasks is crucial. We have defined our task scenario as preliminary diagnosis, which involves making an initial assessment based on recorded dialogues with individuals seeking mental health assistance. This helps physicians quickly grasp key information from these individuals, enhancing diagnostic efficiency. Given the challenges of this task, the models must be highly responsive, capable of accurately understanding and addressing mental health concerns with minimal errors, and they must operate efficiently on limited hardware resources. The key objective is to rigorously benchmark various models and settings, such as Retrieval-Augmented Generation (RAG), to determine the most suitable configurations for these complex research needs.

\subsubsection{Data Collection}

The DAIC-WOZ database \cite{gratch2014distress}, released in 2014, was established with the goal of developing a conversational AI system capable of detecting verbal and nonverbal signs of mental illness. This dataset consists of 187 individual interview sessions, with each session ranging from 7 to 33 minutes and averaging 16 minutes. The data collection process involved not only recording dialogues but also gathering additional information such as assessment scores and visual data.

\begin{figure}[h]
  \centering
  \vspace{10pt}
  \includegraphics[width=0.9\linewidth]{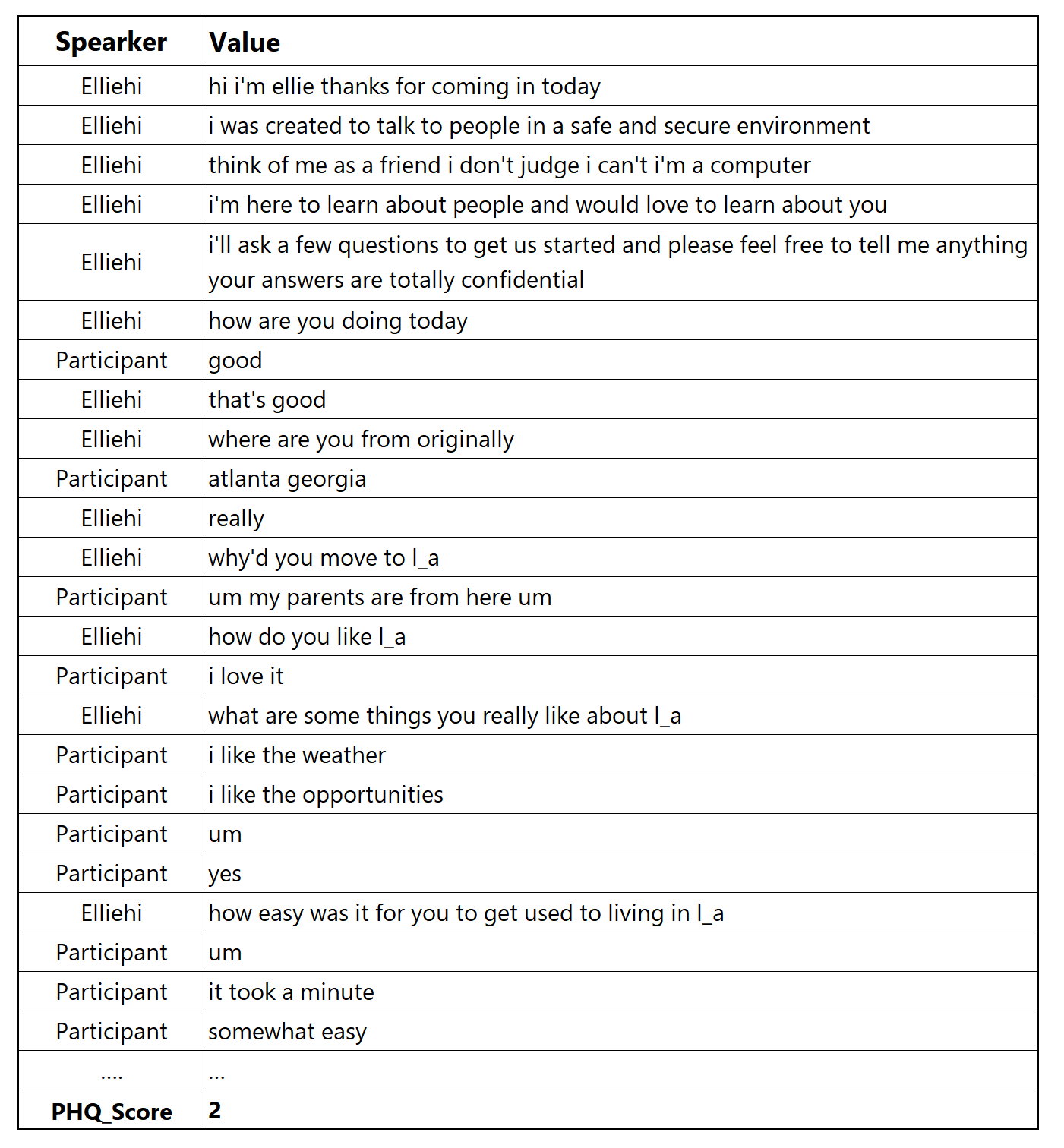}
  \caption{An example from the DAIC-WOZ dataset and the corresponding PHQ-8 score.}
  \label{fig:mental_data}
\end{figure}

The data is organized into a series of folders, each identified by participant IDs ranging from 300 to 492. These folders include a comprehensive array of session materials such as transcripts, audio, and video files. Every session folder is equipped with key data elements: participant IDs, PHQ-8 Binary labels (scores >= 10 indicate potential depression), PHQ-8 Scores, participant gender, and detailed responses to each question of the PHQ-8 questionnaire. This structure provides a comprehensive profile of each participant's mental health status, as illustrated in Figure \ref{fig:mental_data}.

\subsubsection{Data Processing}

Given the rigor and precision of the data collection process, the main steps in data processing primarily involve conversation handling and preparation. Initially, different speakers are labeled as "interviewer," and the ongoing dialogues from the same individual are organized cohesively. Each individual interview record is then structured as a sequence of alternating entries, such as 'interviewer: doctor: interviewer: doctor:…', with a corresponding diagnostic score assigned to each interview.

In the utilization of this data, our input consists of a prompt (a specific question or topic) combined with the interview text and a Retrieval-Augmented Generation (RAG) document. The output is formatted as 'score: x', with an optional explanation to elucidate the reasoning behind the score.

\subsubsection{Experiment Setup}

We employ the CEBench tool to evaluate different model configurations, such as model bits, the application of RAG, RAG document configurations, quantization methods, and chunk sizes. This allows us to assess model performance on the dataset under different conditions, focusing on Mean Absolute Error (MAE) values, specificity, and response times.

\textbf{Hardware Environment}: We assume that the research will be conducted on private servers equipped with NVIDIA A100 GPUs, Intel Xeon Platinum 8380 CPUs, and 1TB of RAM.

\textbf{Benchmark Tool}: CEBench.

\textbf{Model Settings}: We have selected models that can operate on a single A100 GPU, including commonly used open-source models, as listed in Table~\ref{tab:model_summary}. Smaller models offer faster inference times and are preferable for their response times. Larger models have stronger capabilities but operate slower and require significant memory, with the largest model demanding up to 80GB of VRAM.

\begin{table}[ht]
\centering
\renewcommand{\arraystretch}{1.5} 
\setlength{\tabcolsep}{12pt} %
\begin{tabular}{|l|l|l|l|}
\hline
\textbf{Model Name} & \textbf{Parameters} & \textbf{Quantization} & \textbf{Size} \\ \hline
llama2:7b & 6.7B & Q4\_0 & 3.8GB \\ \hline
llama2:7b & 6.7B & Q8\_0 & 7.2GB \\ \hline
llama2:7b & 6.7B & FP16 & 13GB \\ \hline
llama2:13b & 13.0B & Q4\_0 & 7.4GB \\ \hline
llama2:13b & 13.0B & Q8\_0 & 14GB \\ \hline
llama2:13b & 13.0B & FP16 & 26GB \\ \hline
llama2:70b & 69.0B & Q4\_0 & 39GB \\ \hline
llama2:70b & 69.0B & Q8\_0 & 73GB \\ \hline
llama3:8b & 8.0B & Q4\_0 & 4.7GB \\ \hline
llama3:8b & 8.0B & Q8\_0 & 8.5GB \\ \hline
llama3:8b & 8.0B & FP16 & 16GB \\ \hline
llama3:70b & 70.6B & Q4\_0 & 40GB \\ \hline
llama3:70b & 70.6B & Q8\_0 & 75GB \\ \hline
mixtral:8*7b & 46.7B & Q4\_0 & 26GB \\ \hline
mixtral:8*7b & 46.7B & Q8\_0 & 50GB \\ \hline
mixtral:8x22b & 140.6B & Q4\_0 & 80GB \\ \hline
\end{tabular}
\vspace{3pt}
\caption{Parameters of the Selected Models.}
\label{tab:model_summary}
\end{table}

\textbf{Selected Dataset:} DAIC-WOZ database.

\textbf{RAG:} We use DSM-5-TR \cite{DSM5} along with nine other mental health diagnostic manuals as the Retrieval-Augmented Generation (RAG) document source. We experimented with three different quantization options: pq, scalar, and no. Additionally, we varied the chunk size parameters to include 500, 1000, and 2000 and adjusted the top k values at 2, 5, and 10 to observe the impact on the model’s performance in terms of retrieval efficiency and response accuracy.

\textbf{Evaluation Metrics:} MAE, Specificity, Response Time. 

\textbf{Mean Absolute Error (MAE)} is primarily used to assess the model's ability to understand and handle complex tasks related to mental health.  

\textbf{Specificity} (also called True Negative Rate, TNR) measures the accuracy of the pre-diagnosis assistant in determining whether a patient does not have mental health issues.  

\textbf{Response Time} is critical in real-time conversation scenarios as it directly impacts user experience and the usability of the conversational AI.

\subsubsection{Experiment Results}

\begin{figure}[h]
  \centering
  \includegraphics[width=1\linewidth]{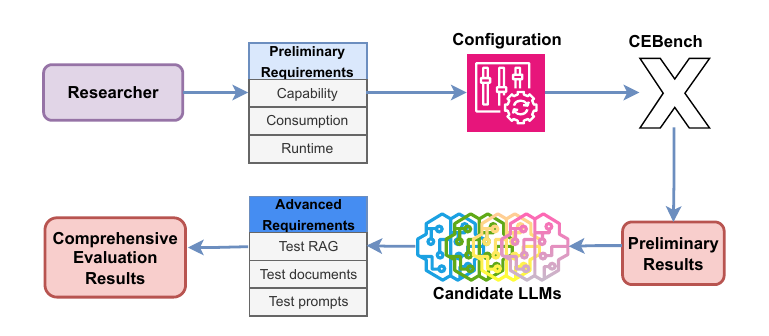}
  \caption{Flowchart of Researchers Using CEBench for Comprehensive Evaluation.}
  \label{fig:mental_uml}
  \vspace{-3mm}
\end{figure}

\textbf{Experimental Procedure} begins with researchers outlining \textbf{Preliminary Requirements} to assess using CEBench, resulting in \textbf{Initial Results}. This step helps filter out models that clearly do not meet the assessment criteria, minimizing evaluation time and resource consumption, and leading to the selection of candidate LLMs. \textbf{Advanced Requirements} are then specified, such as:
\begin{itemize}
\item Testing RAG capabilities
\item Comparing different documents
\item Contrasting various prompts
\end{itemize}
These requirements are submitted to CEBench for further evaluation, ultimately yielding \textbf{Comprehensive Evaluation Results}. Figure \ref{fig:mental_uml} illustrates an example of this process.

\begin{figure}[h!]
    \centering
    \includegraphics[width=0.9\textwidth]{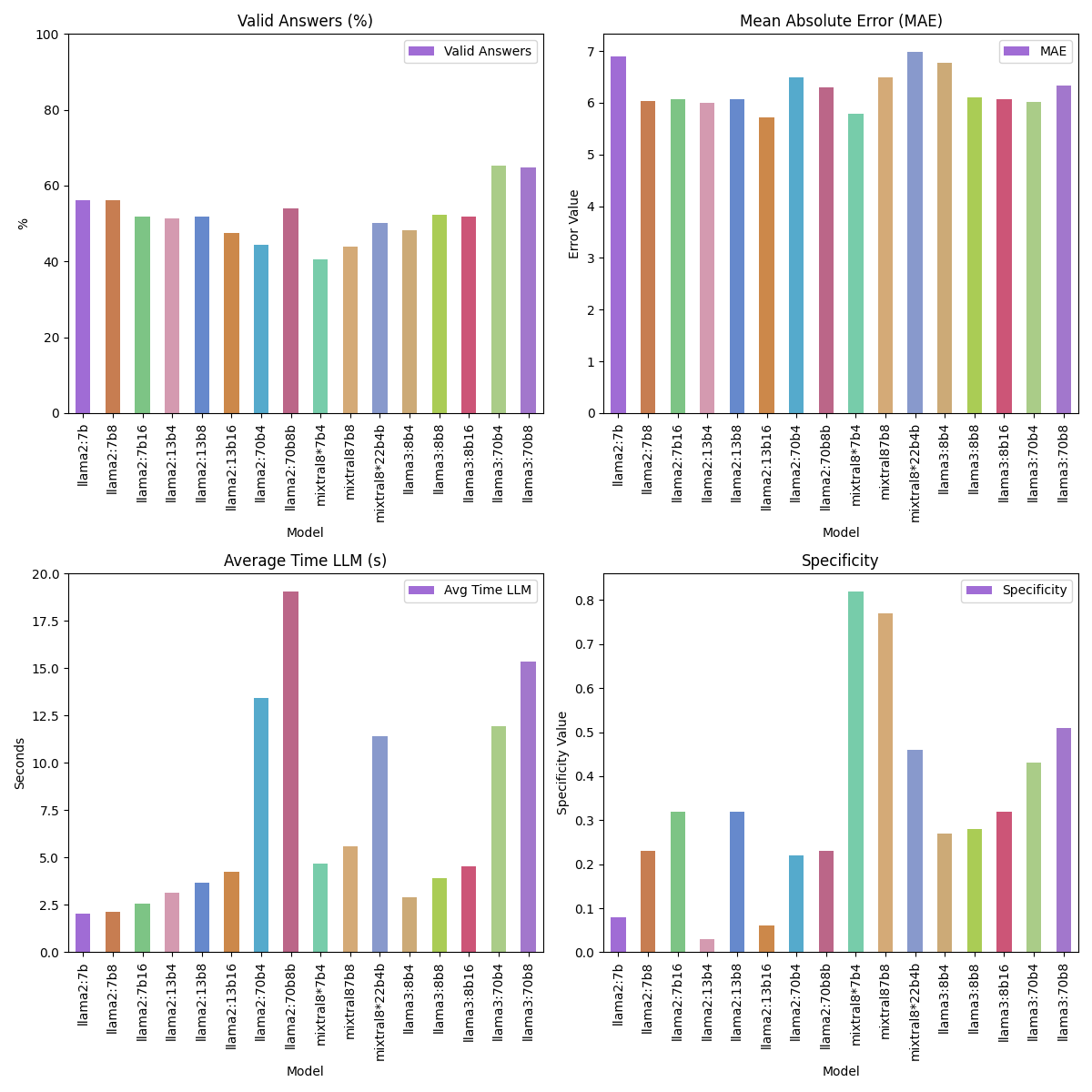}
    \caption{Comparison of performance metrics across different models.}
    \label{fig:Performance Metrics for Various Models}
\end{figure}

\begin{figure}[h]
  \centering
  \vspace{20pt} 
  \includegraphics[width=1\linewidth]{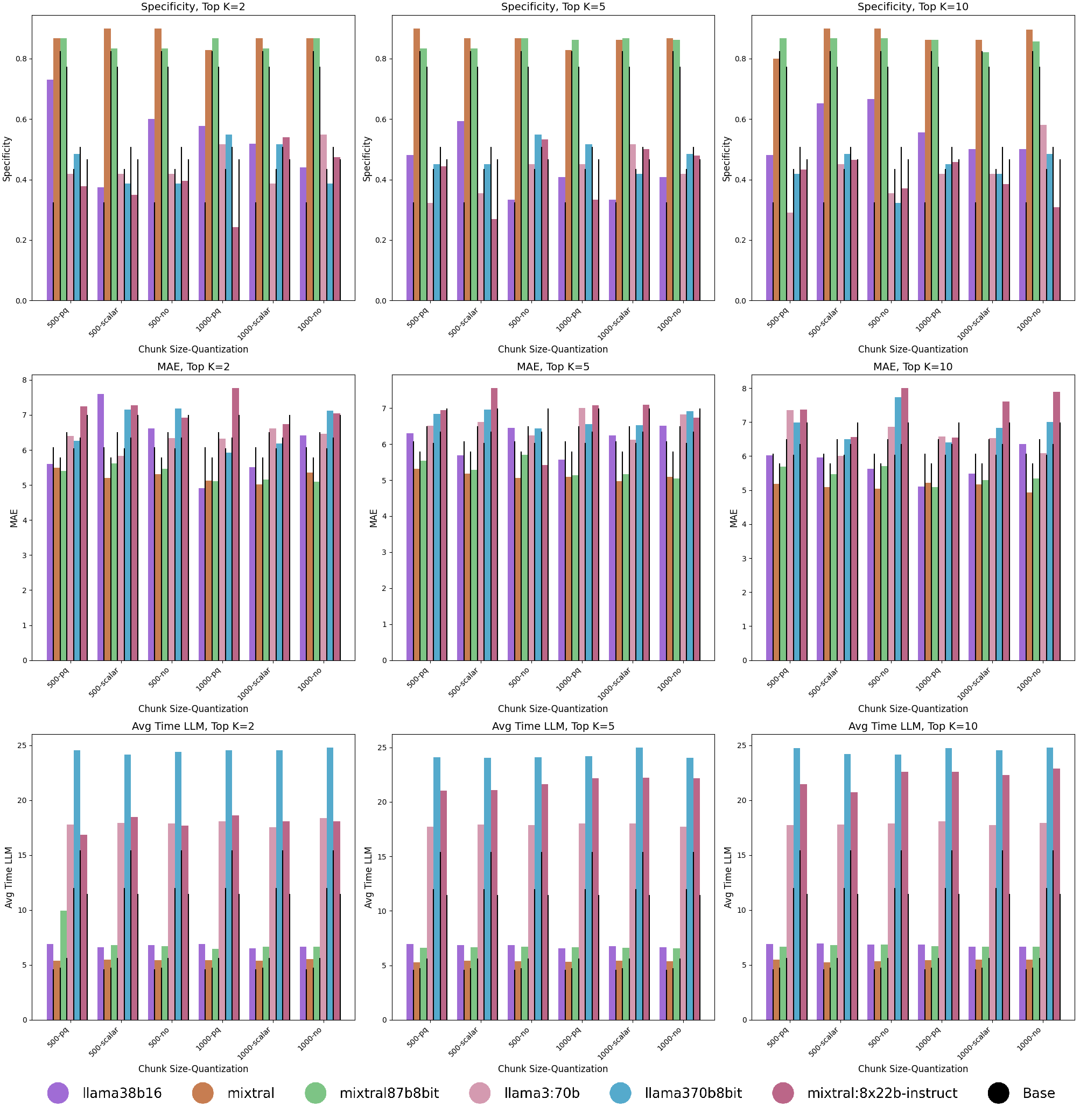}
  \caption{Performance comparison of selected models with and without RAG integration in terms of MAE, runtime, and specificity.}
  \label{fig:mental_rag}
\end{figure}

\textbf{Experimental Results}: We initially utilized CEBench to assess the baseline performance of models for the mental health preliminary diagnosis task. The use of CEBench's logging tool enabled the swift acquisition of results, as detailed in Table \ref{tab:model_performance} and Figure \ref{fig:Performance Metrics for Various Models}. Based on these initial results, we selected 6 well-performing models for further evaluation using CEBench, with the addition of the RAG feature. As shown in Figure \ref{fig:mental_rag}, some models exhibited improvements in MAE and Specificity compared to their base versions after the integration of RAG. However, the runtime of most models increased with the addition of RAG, as the runtime is generally proportional to the input length.

\begin{figure}[h]
  \centering
  \vspace{10pt}
  \includegraphics[width=1\linewidth]{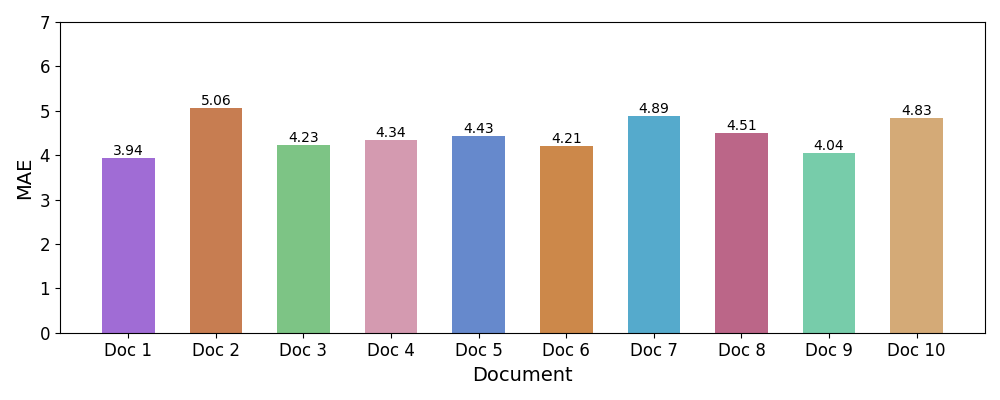}
  \caption{RAG performance across different documents in the Mixtral model.}
  \label{fig:mental_rag_doc}
\end{figure}

We also explored the effects of incorporating RAG documents. In the field of psychological diagnostics, textbooks often guide the identification and treatment of mental health issues, with extensive case studies. We utilized CEBench to investigate the differences in performance between models with and without the use of RAG. Figure \ref{fig:mental_rag_doc} shows the MAE of different documents under the same test questions and model.

Based on these comprehensive results, research institutions can evaluate options and select models tailored to specific needs, such as:

\begin{itemize}
\item For designing a system with an Agent workflow that integrates multiple models for responses, one might consider the fastest model, such as LLaMA3:8b, to ensure swift interaction speeds.

\item For more effective interview summarization, the LLaMA3:70b model could be the preferred choice due to its higher valid answer rate and comparatively high specificity.

\item For pre-diagnostic purposes, the Mixtral:8*7b model could be considered, as it shows the best performance in terms of specificity.

\end{itemize}

\begin{table}[ht]
\centering
\caption{Detailed performance metrics for selected models.}
\label{tab:model_performance}
\begin{tabular}{@{}lcccc@{}}
\toprule
Model                  & Valid Answers & MAE  & Avg Time LLM & Specificity \\ \midrule
llama2:7b 4bit              & 56.15\%       & 6.90 & 2.03         & 0.08        \\
llama2:7b 8bit           & 56.15\%       & 6.03 & 2.13         & 0.23        \\
llama2:7b 16             & 51.87\%       & 6.06 & 2.59         & 0.32        \\
llama2:13b 4bit             & 51.34\%       & 6.00 & 3.17         & 0.03        \\
llama2:13b 8bit          & 51.87\%       & 6.06 & 3.67         & 0.32        \\
llama2:13b 16            & 47.59\%       & 5.72 & 4.23         & 0.06        \\
llama2:70b 4bit             & 44.39\%       & 6.49 & 13.45        & 0.22        \\
llama2:70b 8bit          & 54.01\%       & 6.30 & 19.07        & 0.23        \\
\textbf{Mixtral8*7b 4bit}               & 40.64\%       & 5.78 & 4.69         & 0.82        \\
\textbf{Mixtral8*7b 8bit}         & 43.85\%       & 6.49 & 5.58         & 0.77        \\
\textbf{Mixtral8*22b 4bit} & 50.27\%       & 6.98 & 11.42        & 0.46        \\
llama3:8b 4bit              & 48.13\%       & 6.77 & 2.91         & 0.27        \\
llama3:8b 8bit           & 52.41\%       & 6.10 & 3.91         & 0.28        \\
\textbf{llama3:8b 16}           & 51.87\%       & 6.06 & 4.56         & 0.32        \\
\textbf{llama3:70b 4bit}             & 64.71\%       & 6.02 & 11.94        & 0.43        \\
\textbf{llama3:70b 8bit}          & 65.24\%       & 6.34 & 15.37        & 0.51        \\ \bottomrule
\end{tabular}
\end{table}

\subsubsection{Conclusion}
In this highly specialized task, selecting the appropriate model and settings, including model choice, RAG documents, quantization methods, chunk sizes, search techniques, and top-k settings, is crucial. Analyzing the impact of different configurations on model performance is essential for optimizing the system. The CEBench tool greatly facilitates this process by enabling automated and efficient testing of LLM models' performance under various conditions, without the need for additional coding. This research lays the foundation for developing effective AI-assisted mental health preliminary diagnosis systems, which could significantly improve the efficiency and accessibility of mental health services.

\end{document}